**Article Type:** Original Research Article

**Title:**
Dynamic Modulated Arc Therapy (DMAT): A Time Aware, Modulation Steered Optimization Framework for Next Generation Radiotherapy Delivery

**Running Title:**
Dynamic Modulated Arc Therapy

**Authors:**
Taoran Li; Esa Kuusela; Emmi Ruokokoski; Heini Hyvönen; Jerry Jaboin; Mirko Myllykoski; Jussi Nurminen; Riku Paananen; Jarkko Peltola; Marko Rusanen; Martin Sabel; Kevin Moore; Christopher Boylan

**Affiliation:**
Varian Medical Systems Inc., a Siemens Healthineers company, Palo Alto, CA, USA

**Corresponding Author:**
Taoran Li, PhD
Varian Medical Systems Inc., a Siemens Healthineers company
Palo Alto, CA, USA
Email: taoran.li@siemens-healthineers.com

# Dynamic Modulated Arc Therapy (DMAT): A Time-Aware, Modulation-Steered Optimization Framework for Next-Generation Radiotherapy Delivery

## Abstract

**Background:** Conventional VMAT optimization largely treats delivery time and deliverability as emergent properties of simplified, control-point-centric models that often ignore finite acceleration and other dynamic limits. As modern linacs continue to increase maximum axis speeds and dose rates, there is a growing need for a planning paradigm that makes the plan quality-time trade-off explicit and steerable, especially when shorter treatments can provide clinical benefits beyond throughput.

**Purpose:** This work introduces Dynamic Modulated Arc Therapy (DMAT), a time-aware, modulation-steered framework that jointly optimizes dosimetric quality, delivery time, and modulation complexity. Here, time-aware denotes that delivery time is computed within the optimizer from an explicit model of machine dynamics, including finite acceleration and axis synchronization, rather than inferred after optimization; modulation-steered denotes that leaf-travel allowance, control-point density, total monitor units, and aperture complexity are governed together by a single user-selected modulation level, making the amount and distribution of modulation an explicit input to the optimizer. Clinical goals enter the optimizer as metric-based cost functions defined by each criterion's metric, goal value, acceptable variation, and priority, with supporting objectives generated automatically.

**Methods:** DMAT couples (1) direct machine emulation that accounts for axis synchronization and dynamic limits, (2) dynamic modulation control, and (3) clinical metrics as optimization cost functions. A user-selected modulation level (−3 to +3) governs leaf-travel allowance, total MU behavior, aperture-complexity penalties, and control-point (CP) density. Plans are created by defining a treatment trajectory, initializing CP geometry, leaf positions, and MU, then iteratively improving the solution with a progressive-resolution loop alternating dosimetric updates with sequencing/deliverability updates, followed by post-processing to reduce complexity with minimal dose change. CP density is adapted nonuniformly: a preliminary uniform-CP optimization is followed by redistribution according to combined delivery-complexity and geometric information. DMAT was evaluated using a hypothetical accelerated C-arm delivery system (max gantry speed 2.5 RPM, max MLC speed 6.25 cm/s, max dose rate 3000 MU/min) on representative head-and-neck, lung SBRT, and prostate SBRT cases.

**Results:** Across all cases, increasing modulation level produced plans with increased modulation surrogates (e.g., MU/Gy and aperture complexity) and progressively longer delivery time, while adaptive non-uniform CP allocation concentrated additional CPs in arc sectors where higher angular resolution was most beneficial. The resulting quality–time trade-off space was disease site dependent: the head-and-neck case showed substantial plan-quality gains with increased modulation and CP density, whereas prostate SBRT and lung SBRT exhibited smaller incremental quality improvements beyond baseline despite longer delivery times. When efficiency was prioritized (negative modulation levels), DMAT predictably reduced modulation and maintained a constant CP budget while shortening delivery time, producing corresponding and quantifiable reductions in plan quality.

**Conclusions:** DMAT establishes a time-aware, modulation-steered planning and delivery methodology in which plan quality and modulation complexity are co-optimized under explicit clinical and user control using machine-aware timing. Accurate delivery time is exposed to users during planning through advanced machine emulation. By doing so, DMAT makes quality–time trade-offs transparent, predictable, and navigable, providing a practical foundation for leveraging next-generation delivery systems' capabilities and supporting time-constrained workflows such as motion-sensitive treatments and adaptive radiotherapy.

## Introduction[1]

### *Background and Motivation*

The evolution of arc therapy has been marked by a steady expansion of dynamic capabilities, beginning with early innovations such as Yu's IMAT (1), which introduced continuous gantry rotation with modulating MLC apertures and laid the groundwork for intensity-modulated arc delivery. Around the same time, TomoTherapy, introduced by Mackie et al (2,3), combined rotational beam delivery with helical couch motion and integrated imaging, establishing a CT-like paradigm for highly conformal, image-guided treatments (4). The emergence of VMAT brought broader adoption of aperture-based optimization for improved delivery efficiency, enabling IMRT-quality dose distributions to be delivered in a single arc through coordinated modulation of dose rate, gantry speed, and MLC positions (5–9). These techniques achieved widespread clinical adoption by striking a favorable balance between plan quality and delivery efficiency.

Building on this foundation, subsequent developments have focused on expanding the geometric degrees of freedom available to arc therapy. Dynamic collimator and couch rotation for VMAT were first proposed by Yang et al (10), predating the noncoplanar optimization strategies that followed. Noncoplanar optimization strategies most notably 4π planning and commercial solutions such as HyperArc broadened the orientation space and automated couch-gantry coordination to achieve sharper dose gradients and improved sparing of organs at risk (OARs), particularly in stereotactic radiosurgery (SRS) (11–14). More recently, trajectory-modulated VMAT, dynamic couch rotation (DCR-VMAT), and dynamic collimator rotation (DC-VMAT) have demonstrated additional dosimetric gains by co-optimizing three-dimensional beam trajectories and collimator orientations on modern linear accelerators

[1] All trademarks are the property of their respective owners.

(15–18). Parallel advances, including non-uniform control-point (CP) density(19), jaw tracking, and multicriteria optimization (MCO), have further refined dosimetry and enhanced the planner's ability to explore trade-offs between competing clinical goals (20–23).

Despite this continued expansion of dynamic and geometric capabilities, the fundamental formulation of commercial VMAT optimization has remained largely unchanged. The optimization problem is still centered on a predefined set of control points, with MLC leaf positions and delivered monitor units (MUs) as the primary decision variables (5–8). Even in advanced workflows that incorporate noncoplanar beam trajectories, couch and collimator rotation, or optimized isocenter placement, these geometric decisions — that is, the trajectory itself and the number and angular spacing of the control points sampled along it — are typically fixed in a separate, preceding step and then held constant during the subsequent optimization of MLC leaf motion and monitor units. The degrees of freedom exposed to the optimizer are therefore restricted a priori to leaf positions and monitor units at a predetermined geometry, so trajectory design and aperture modulation are optimized sequentially rather than jointly within current planning frameworks.

An additional limitation arises from the representation of treatment plans within the optimizer. VMAT plans are parameterized by DICOM control points, which act as geometric waypoints along the delivery path. Each control point specifies gantry angle, leaf positions, and cumulative MU, but does not explicitly encode temporal information (8). The task of translating these waypoints into a time-resolved delivery that satisfies machine constraints on velocity, acceleration, deceleration, and higher-order motion dynamics such as jerk is deferred to the treatment control system. Consequently, most VMAT optimizers consider only maximum axis speeds (6,24) and implicitly assume instantaneous acceleration. In practice, acceleration limits may require the machine to slow down or halt during delivery, leading to longer treatment times than predicted and producing discrepancies between the optimized and delivered trajectories. This effect is most pronounced when delivery would otherwise proceed near the maximum gantry speed, and is correspondingly less relevant when the maximum dose rate is the limiting factor and gantry motion is already slowed.

Because control points define the degrees of freedom available to the optimizer, both their number and spatial distribution are typically fixed prior to optimization (1,5,6,25). In these formulations the continuous delivery arc is approximated, for optimization tractability, by a series of evenly-spaced static control points; because these control points are the only degrees of freedom exposed to the optimizer, this discretization itself constrains the achievable modulation independent of the machine's mechanical capabilities. To ensure deliverability and computational efficiency, further simplifying assumptions are often imposed. For example, in conventional VMAT planning, control points are prepopulated at approximately uniform gantry-angle spacing, gantry positions remain fixed during optimization and inter-control-point leaf motion is constrained such that required leaf travel does not exceed the time required for gantry rotation. While these assumptions simplify the optimization problem, they also restrict the achievable modulation. In particular, limiting leaf travel per unit gantry angle can under-modulate large or geometrically complex targets, and fixing the number of control points restricts the locations at which leaf velocities can change, leading to coarse and potentially suboptimal leaf trajectories. Together, these constraints both reduce the optimizer's ability to leverage machine-level dynamic capabilities to improve plan quality and often require the planner to utilize multiple arcs to obtain sufficient control of the resultant dose distribution (1,5,6).

In parallel, the formulation of the optimization objectives introduces further restrictions. For example, Varian VMAT optimizers rely on quadratic dose-volume histogram (DVH)-based objective functions, which serve as numerical surrogates for clinically defined treatment goals and constraints. Clinical intent is used here to mean the prioritized set of clinical criteria — typically dose–volume criteria such as D95% > 95% or a maximum organ-at-risk dose — together with their relative priority and the degradation regarded as acceptable when they compete. Although the Varian Ethos Intelligent Optimizer Engine (26,27) seeks to bridge this gap through a dynamic conversion layer that maps clinical intent onto DVH-based objectives, this separation introduces inefficiencies. Specifically, the conversion layer evaluates plan quality using a different representation than the one guiding the optimizer's numerical search. As a result, the optimization process may be driven by surrogate cost functions that only partially reflect the planner's true clinical goals and priorities.

Meanwhile, conventional radiation therapy hardware has continued to advance. Maximum gantry and MLC leaf velocities have increased substantially, along with the available dose rates; for example, the Varian Halcyon system can achieve gantry rotation speeds of up to 24 deg/s (28). However, the maximum acceleration and deceleration capabilities of mechanical axes have not scaled proportionally. As a result, acceleration limits play an increasingly dominant role in determining achievable delivery speed. If not explicitly accounted for, part of the nominal gain in maximum velocity may be lost to acceleration and deceleration phases, or to the requirement to smooth velocity transitions at control-point boundaries, the extent of which is machine- and plan-specific. At the same time, higher achievable speeds have renewed interest in treatment-time optimization: in some scenarios, faster gantry motion could enable delivery of entire treatments within a single breath-hold, while in others it permits more complex treatments to be delivered within clinically acceptable time frames (29,30). Other relevant developments include continued improvements in dose calculation accuracy and modern high-end GPUs that allow optimization and dose calculation to be performed rapidly even for large, highly modulated patient cases.

Taken together, these limitations and opportunities suggest that, although modern VMAT systems increasingly exploit complex beam trajectories and advanced hardware capabilities, the underlying optimization framework remains constrained by control-point–centric representations (that is, formulations in which the continuous arc is represented as a fixed, predefined set of static control points and only the MLC leaf positions and MU at those points are optimized, rather than the trajectory geometry, the control-point distribution, or the delivery timing itself), implicit timing assumptions, and rigid objective formulations.

Consequently, the full potential of enhanced machine capabilities—specifically the higher maximum gantry and MLC leaf speeds and elevated dose rates of modern linacs, together with the finite acceleration/deceleration headroom that determines how much of that nominal speed can actually be exploited during delivery—cannot be realized within the existing paradigm. Addressing these challenges requires a reformulation of the optimization problem that explicitly incorporates trajectory parameterization, temporal dynamics, and clinical intent. This motivation underlies Dynamic Modulated Arc Therapy (DMAT), a framework in which delivery complexity and clinical quality are treated as jointly optimized, user-steerable variables, with machine motion and timing embedded directly in the optimizer so that plan quality, delivery duration, and mechanical feasibility are balanced throughout the numerical search.

## Methods and Materials

### *Overview of the DMAT Concept*

To improve planning efficiency and clinical relevance, the optimization framework should better reflect how planners communicate treatment intent to the optimizer. This includes not only the relative importance assigned to competing clinical goals, but also explicit control over the trade-off between plan complexity, delivery time, and dosimetric quality.

As the dynamic capabilities of treatment machines continue to advance, it becomes increasingly important to ensure consistency between the optimizer's assumptions and the physical behavior of the delivery system. In practice, increases in maximum axis velocities—particularly for the gantry and MLC leaves—are often easier to achieve than corresponding gains in maximum acceleration or deceleration. Gantry acceleration, in particular, requires substantial mechanical power, making it costly to increase. Consequently, any need to slow gantry motion to accommodate other axes may trigger additional deceleration–acceleration cycles, leading to a disproportionate increase in delivery time relative to the required adjustment itself. Because inter-control-point leaf displacement is separately constrained, variation in monitor units per degree along the arc becomes a principal source of changing synchronization requirements between dose rate and gantry motion; whether such variation actually requires gantry slowing depends on the available dose-rate and gantry-speed headroom, and it is addressed by the MU smoothing and MU-uniformity terms of the modulation cost described below. This effect becomes more pronounced as control-point density increases and time intervals between successive control points shorten, thereby increasing sensitivity to finite acceleration limits. These considerations motivate an optimization framework in which leaf positions and MUs are not treated independently at individual control points but are instead coordinated through an explicit temporal and mechanical model.

DMAT is a new arc plan generation framework that addresses these limitations by treating modulation complexity as an explicit, co-optimized resource alongside dosimetric objectives. DMAT embeds a physically informed model of machine motion and timing directly into the optimizer so that plan quality, delivery duration, and mechanical feasibility are balanced during the numerical solution search. Control points are distributed non-uniformly based on preliminary modulation and geometric complexity, concentrating degrees of freedom where they are most effective and avoiding over-parameterization of low-impact segments. This enables DMAT to allocate modulation where it yields the greatest clinical benefit while avoiding modulation patterns that would be costly or infeasible to deliver.

DMAT is built on three coupled design principles embedded in a single optimization program:

1. **Direct machine emulation**, which models delivery timing and mechanical feasibility.
2. **Dynamic modulation control**, which regulates how modulation is distributed across the arc.
3. **Clinical metrics as cost functions**, which align optimization directly with clinically meaningful planning goals.

These pillars interact bidirectionally: complexity influences time through mechanical limits; time bounds achievable gradients; and both mediate how efficiently clinical metrics can be attained. The formulation exposes these couplings so planners can steer trade-offs explicitly.

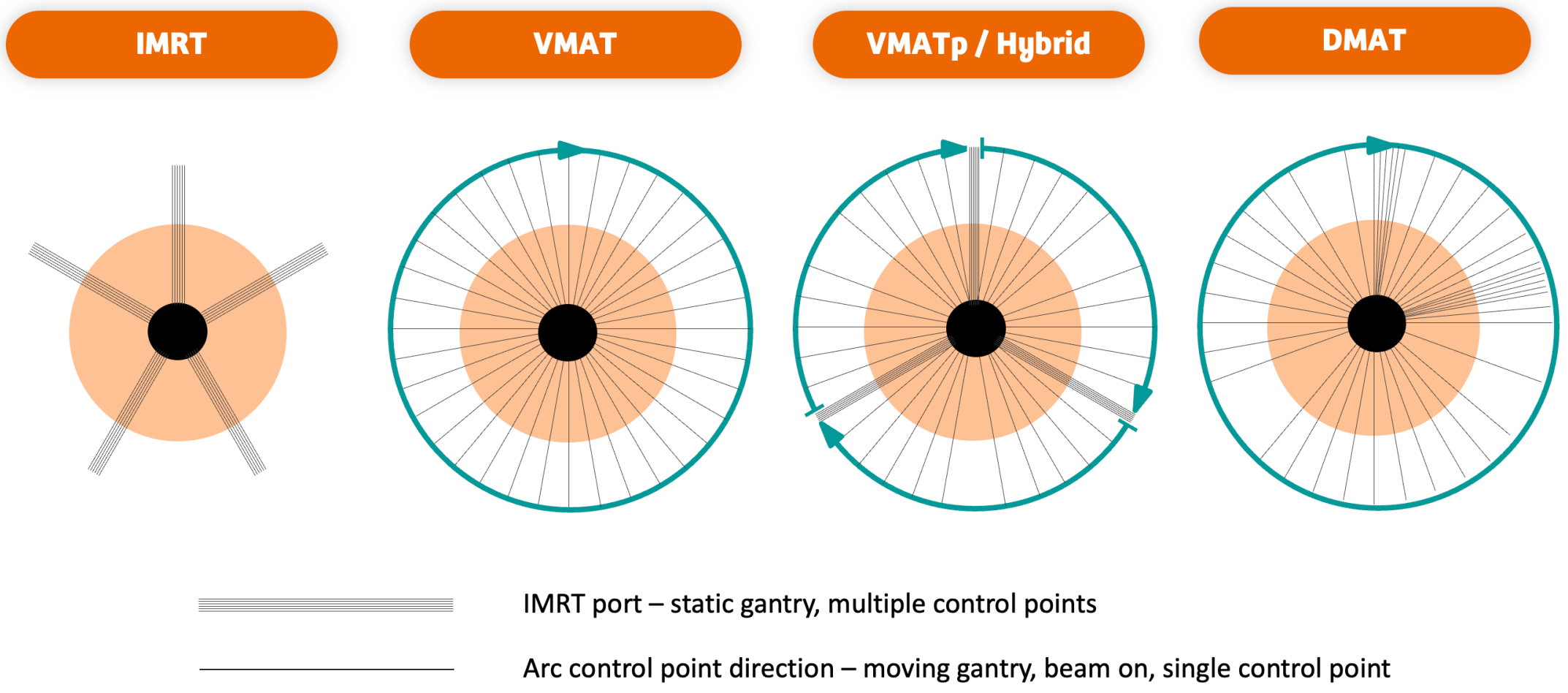

Figure 1 Conceptual comparison of IMRT, VMAT, VMATp (RapidArc Dynamic), and DMAT. In this illustration, line density denotes the relative concentration of delivery resources across beam directions or arc sectors: sparse elements indicate lower local modulation density, while dense elements indicate sectors receiving greater modulation emphasis and/or higher effective angular resolution. DMAT extends arc therapy by allowing this modulation allocation to be distributed non-uniformly according to anatomy, clinical objectives, and delivery-time constraints, thereby making the quality–time trade-off explicit. (Note that dose rate modulation and collimator rotation are not shown.)

### *Machine emulation*

Conventional VMAT optimization often approximates machine delivery using only maximum axis velocities. For example, the duration of a control point interval may be determined by the axis requiring the longest time to complete its motion, with all other axes scaled to move synchronously.

In practice, however, the treatment machine must translate the sequence of discrete control points into continuous, time-resolved mechanical trajectories. This process generally requires that the axes do not operate at the velocities defined by the simplified delivery timing model at all times but instead adjust their motion to ensure smooth synchronization across axes while respecting constraints on acceleration, deceleration, and higher-order dynamics. Consequently, accurate modeling of delivery requires that the optimizer captures essential aspects of machine trajectory generation and explicitly accounts for the acceleration limits of individual axes. Within DMAT, this is achieved by computing, for each axis—including the gantry, collimator, and other (potentially non-optimized) degrees of freedom—the maximum allowable velocity changes that are consistent with these dynamic constraints.

A direct consequence of this more detailed machine emulation is the ability to generate reliable predictions of treatment delivery time. Velocity-only models fail to capture the effects of finite acceleration and ignore the need for temporal smoothing of velocity changes at control-point locations. As the magnitude of required velocity changes increases relative to permissible acceleration and deceleration limits, these simplified models increasingly underestimate delivery time. By explicitly incorporating axis-specific dynamic constraints, DMAT avoids such discrepancies and provides delivery-time estimates that remain accurate even for highly modulated or dynamically complex treatment plans. Reliable delivery-time prediction is therefore not a separate objective but an observable consequence of enforcing these dynamic constraints during optimization: a trajectory that respects the axis limits is, by construction, one whose predicted duration matches its delivery.

## *Dynamic Modulation control*

Dynamic modulation control allows the optimizer to regulate both the amount and spatial distribution of modulation. In DMAT, modulation is decomposed into four independently controlled components: leaf travel allowance, control-point density, total monitor units, and aperture complexity.

Leaf travel allowance determines how far MLC leaves may move between neighboring control points. Increasing this allowance gives the optimizer more freedom to create rapidly changing apertures, enabling a higher number of distinct dose levels to be delivered. However, if leaf motion becomes the rate-limiting axis, it will also increase delivery time. DMAT addresses this trade-off by permitting any leaf motion that does not extend delivery time, while applying an additional modulation-dependent allowance when greater aperture complexity is requested. Within this scheme the optimizer preferentially maintains consistent gantry motion and achieves the requested modulation through the other axes—MLC leaf motion and, at higher modulation levels, greater dose-rate variation—first; gantry-speed modulation (slowing the gantry) is invoked only when the requested modulation cannot otherwise be accommodated within the nominal gantry motion, with the extent to which each axis is exploited depending on the specific case.

Control point density determines the temporal resolution of the treatment plan. Higher control point density allows more detailed leaf trajectories and finer angular sampling within each sector of the progressive-resolution optimization, so that the fluence delivered across a sector can more closely match the fluence scored for it (see Optimization Procedure). However, when CP spacing becomes sufficiently fine, the corresponding reduction in inter-CP time intervals renders machine dynamic limits—such as acceleration and jerk—increasingly relevant. Excessive control-point density may therefore compromise deliverability despite increased modulation freedom. DMAT addresses this by assigning control points non-uniformly. A preliminary optimization pass is first performed using an angularly uniform control point distribution. The preliminary control-point sequence is analyzed to estimate local delivery-complexity demand from its aperture and meterset characteristics. This information is combined with geometric information derived from target and organ-at-risk projections, and control points are redistributed toward arc sectors where greater angular resolution is expected to be most useful. The total control-point budget increases with modulation level and the aggregate complexity of the preliminary solution. Figure 1 shows a schematic presentation of the DMAT control point positioning and a comparison with prior treatment modalities.

Total monitor units provide another measure of modulation. Although higher MU is often associated with smaller, more modulated apertures, a high MU does not necessarily imply small aperture sizes, since a relatively large aperture opening can also carry high MU; the determinant of dose-calculation and delivery robustness is therefore aperture complexity rather than total MU per se. Small apertures are also assumed to produce less robust plans, where minor dose calculation inaccuracies and deviations between planned and actual leaf positions have an amplified effect relative to total dose delivered, making machine QA increasingly challenging. DMAT addresses both concerns by controlling total MU level and aperture complexity. It defines an expected MU range using a complexity-aware initialization scheme and discourages excessive drift from this range during optimization, while directly penalizing aperture complexity through an objective that minimizes the aperture's penumbra ratio, defined here as the ratio of the aperture perimeter (the penumbra-generating edge length) to the enclosed aperture area, such that more irregular or fragmented apertures incur a higher penalty.

In this study, these modulation controls were governed by a single user-defined modulation level ranging from −3 to +3. A value of 0 corresponds approximately to conventional VMAT-like modulation. Negative values emphasize shorter delivery time, lower complexity, and greater robustness. Positive values allow increased leaf travel, higher control point density, more complex apertures and higher MU.

Level-0 VMAT-like fixed-CP reference

Because the accelerated delivery system is a research construct that is not implemented in a clinical treatment planning system, the commercial Photon Optimizer cannot generate a VMAT plan for this machine. All plans were therefore generated using the same prototype DMAT optimizer and machine model. Modulation level 0 was defined a priori as a VMAT-like fixed-control-point reference for comparison with the other modulation levels; it is not intended to reproduce commercial VMAT or support a claim of superiority over VMAT on current clinical systems. Level 0 used uniform 2° control-point spacing, a fixed control-point budget of 180 CPs in a full arc, conventional aperture and MU modulation constraints, no adaptive control-point-density increase, and no efficiency-biased delivery-time minimization. Negative levels imposed stronger efficiency and complexity constraints, whereas positive levels relaxed these constraints and enabled adaptive, non-uniform CP redistribution.

### *Clinical Metrics as Cost Functions*

DMAT uses clinical metrics directly in the dosimetric cost function rather than relying only on voxel-wise quadratic penalties. This allows the optimizer to represent treatment intent in terms familiar to planners, such as target coverage, organ-at-risk dose limits, mean dose, maximum dose, dose-at-volume, volume-at-dose, and gEUD.

The dosimetric cost is written conceptually as:

$$C_D(\boldsymbol{m}) = \sum_i f_i(m_i),$$

where $m_i$ is a dosimetric metric (a scalar value function of the dose distribution, such as “mean dose of left parotid gland” or “maximum dose of spine”) and $f_i$ is a continuous, monotonic, and convex function composed of linear segments joined at smoothed inflection points. The locations of the inflection points are determined by the clinical goal values and their acceptable variation ranges, while the slopes are assigned based on the relative priorities of the corresponding clinical goals. This cost function formalism supports a single static cost function to describe the various trade-offs in optimization depending on the values of the metrics and the level of satisfaction of the prioritized goals.

The distinction intended by "clinical metrics as cost functions" is not that quantities such as Dx, Vx, or mean dose cannot be specified as objectives in existing treatment-planning systems, but that each criterion is retained as a complete specification — metric, goal value, acceptable-variation range, and priority — which directly defines its cost function and remains available to the optimizer throughout the search. The criterion also specifies the desired result rather than the numerical means of producing it: the several supporting objectives or helper structures typically needed to achieve one criterion are generated automatically from the prioritized goal list and the patient geometry rather than being constructed and re-tuned by the planner. Where a criterion is instead converted up front into quadratic dose-based objectives, the quantity minimized becomes the weighted sum of squared voxel-dose deviations, so the optimizer no longer represents by how much a criterion is missed; the influence of each objective then depends on its weight together with structure volume and voxel count, so equal weights do not correspond to equal clinical importance; and the acceptable degradation is not recoverable from the converted objectives, leaving the compromise among conflicting goals dependent on the weighting chosen by the planner. Retaining the metric, goal, acceptable range, and priority as the optimization specification keeps the cost expressed in the same quantities used to evaluate the plan, which is also what allows plans generated at different modulation levels to be compared on identical clinical terms.

The supported metric types include:

1. **Volume-at-dose**, such as $V_{XGy}$ or $V_{X\%}$, representing the volume of a structure receiving at least a specified dose.
2. **Dose-at-volume**, such as $D_{Xcc}$ or $D_{X\%}$, representing the dose received by a specified volume of a structure. Minimum and maximum dose can be represented as limiting cases of this metric.
3. **Generalized equivalent uniform dose (gEUD)** (31,32), which summarizes heterogeneous dose distributions using a structure-specific volume-effect parameter. Mean dose can be represented as special case of this formulation.

## *DMAT plan creation*

DMAT plan creation begins with definition of the treatment trajectory. The trajectory specifies gantry angle, collimator angle, and any additional degrees of freedom, such as couch angle or isocenter position, as functions of a monotonically increasing trajectory point. These trajectory points may be generated from user inputs, such as gantry start and stop angles, or obtained from a separate trajectory-optimization stage.

A set of control points is then placed along the trajectory. Each control point contains the interpolated machine geometry and the optimization variables, including MLC leaf positions and monitor units. The control point spacing may be uniform initially, but DMAT can redistribute control points based on the modulation and geometric complexity of the preliminary solution.

The nominal optimization problem is formulated as

$$\underset{\{z_k\}}{\operatorname{argmin}}\, C(\{x_k\}) \quad ,$$

where $C(\{x_k\})$ is the total cost function evaluated over all control points $\{x_k\}$ including also the trajectory degrees-of-freedom. The optimization variables $\{z_k\}$ contain leaf positions at control points and delivery MUs of the previous control point interval. The optimization problem definition includes any applicable constraints on leaf positions or other deliverability requirements.

## *Total Cost function*

The total cost function in the DMAT framework contains two main components:

$$C = C_D + C_Z$$

where $C_D$ is the dosimetric cost based on clinical metrics, and $C_Z$ is the modulation and deliverability cost.

The modulation cost includes penalties related to leaf motion, scaffold constraints, aperture complexity, leaf acceleration, total MU, MU smoothing, and MU uniformity. It penalizes plan characteristics that reduce deliverability or clinical efficiency. It discourages excessive leaf motion between control points, mechanically valid but clinically inefficient leaf configurations, overly irregular apertures, and motion patterns requiring large velocity changes that would trigger delivery slowdowns. It also maintains clinically reasonable total MU values and controls abrupt or highly non-uniform MU distributions across the arc. These characteristics are quantified explicitly: excessive leaf motion as the summed inter-control-point leaf displacement exceeding the deliverable allowance (cm); leaf acceleration as the squared velocity change between successive intervals ([cm/s$^2$]); aperture irregularity as the perimeter-to-area (penumbra) ratio defined above; and MU non-uniformity as the variation of MU per degree across the arc. Each term is weighted by a modulation-level–dependent factor.

Each modulation term has an associated weight, allowing the optimization behavior to be adjusted according to the selected modulation level and clinical priorities.

### *Optimization Procedure*

DMAT is implemented as an iterative improvement procedure rather than a global optimization method. The procedure consists of five main steps:

1. Create the treatment trajectory.
2. Initialize control points, leaf positions, and MU values.
3. Conduct a preliminary iterative optimization pass, analyze the obtained sequence for complexity and modulation and redistribute control points accordingly.
4. Iteratively improve the solution using dosimetric and sequencing updates.
5. Post-process the plan to improve deliverability while preserving dose quality.

Initialization establishes a starting solution whose overall modulation approximately corresponds to the user-selected modulation level; it does not define the final optimized apertures. For low- to mid- modulation levels, the initial apertures are generated from the beam's-eye-view (BEV) projection of the target, with the leaves opened to the target outline and then moved inward by an amount that increases with the requested modulation. For the higher positive levels, an initial discrete-angle, IMRT-like fluence optimization is instead performed once, prior to the first progressive-resolution level, and used to generate a more highly modulated, anatomy-informed sliding-window seeding aperture sequence for the main optimization. In both cases the control points are then placed at uniform angular spacing and a preliminary aperture-based optimization is performed, after which the resulting sequence is analyzed to determine the nonuniform control-point distribution. The modulation level also affects the number and distribution of control points.

During optimization, DMAT uses a progressive-resolution strategy (5) in which the optimization variables are the MLC leaf positions and segment monitor units. At each resolution level, the control points falling within a given arc length are grouped and their combined delivery is represented as a single sector fluence that is used to score the dosimetric cost and guide the aperture and MU updates; the resolution level defines this grouping arc length and is progressively reduced over the course of the optimization. Each dosimetric update is followed by leaf-position and MU adjustments that minimize the sequencing and deliverability costs while keeping the aperture-generated fluence close to the scored sector fluence, so the optimizer improves clinical metrics while maintaining feasible, directly deliverable machine trajectories.

DMAT is thus fundamentally a direct-aperture optimization method (33): the optimizer does not first generate a final ideal (beamlet) fluence map that is subsequently leaf-sequenced into the delivered control-point sequence. The only exception is the higher-modulation initialization described above, where a brief discrete-angle, IMRT-like fluence optimization is used solely to seed the sliding-window aperture sequence before the direct-aperture optimization process, rather than as a separate beamlet-optimization and leaf-sequencing workflow.

After the iterative optimization loop, post-processing reduces delivery slowdowns caused by large leaf-motion and gantry-velocity changes between neighboring control points. Closed MLC leaf-pair positions are adjusted to reduce peak leaf velocities while remaining within machine constraints and field boundaries. Minor adjustments are also made to control-point gantry positions to reduce changes in gantry velocity, while preserving the original gantry-angle order and limiting deviations from the initial positions.

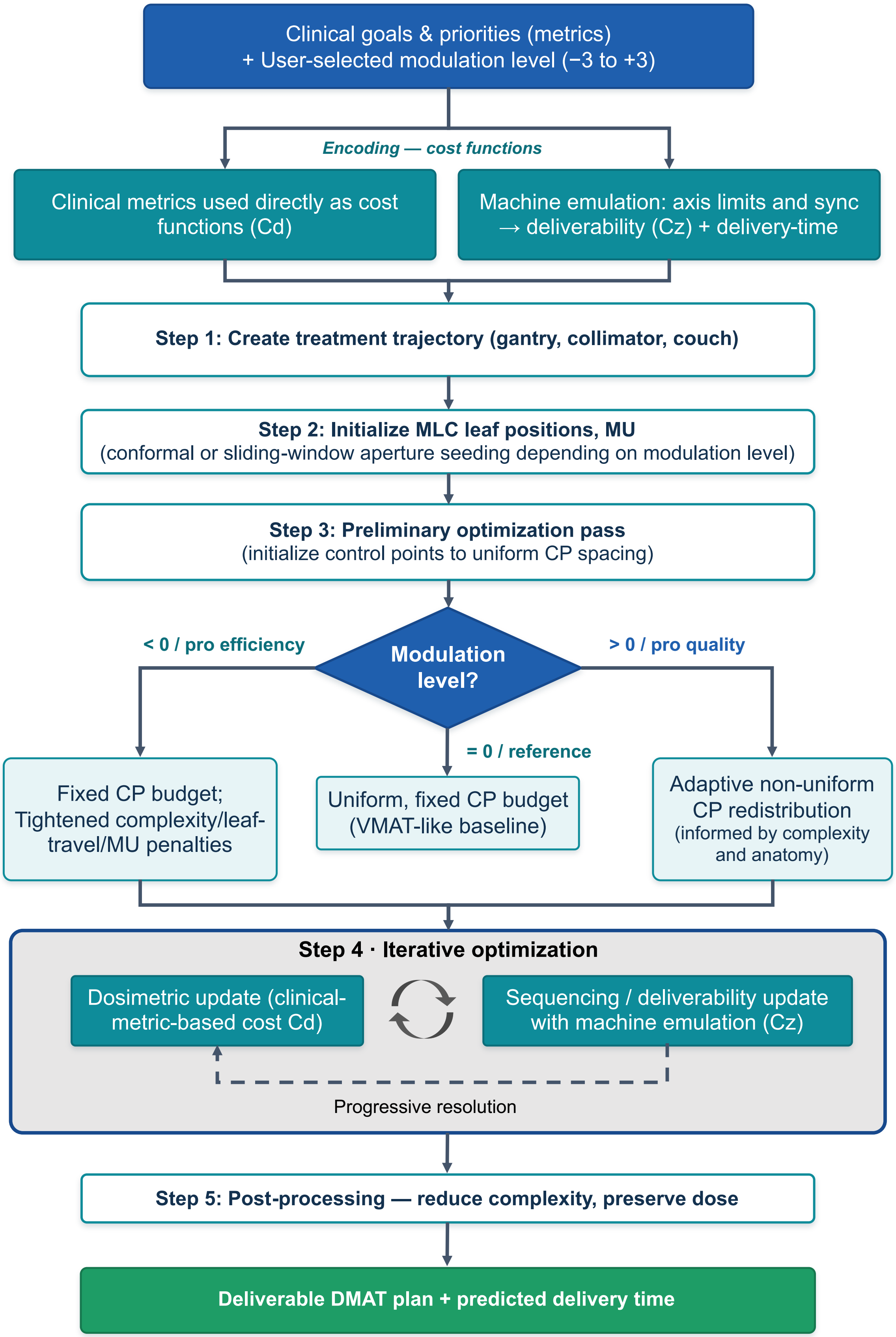
Clinical goals & priorities (metrics)
+ User-selected modulation level (−3 to +3)
Encoding — cost functions
Clinical metrics used directly as cost functions (Cd)
Machine emulation: axis limits and sync → deliverability (Cz) + delivery-time
Step 1: Create treatment trajectory (gantry, collimator, couch)
Step 2: Initialize MLC leaf positions, MU
(conformal or sliding-window aperture seeding depending on modulation level)
Step 3: Preliminary optimization pass
(initialize control points to uniform CP spacing)
Modulation level?
< 0 / pro efficiency
> 0 / pro quality
= 0 / reference
Fixed CP budget; Tightened complexity/leaf-travel/MU penalties
Uniform, fixed CP budget (VMAT-like baseline)
Adaptive non-uniform CP redistribution
(informed by complexity and anatomy)
Step 4 · Iterative optimization
Dosimetric update (clinical-metric-based cost Cd)
Sequencing / deliverability update with machine emulation (Cz)
Progressive resolution
Step 5: Post-processing — reduce complexity, preserve dose
Deliverable DMAT plan + predicted delivery time

Figure 2 DMAT optimization workflow, from encoding of clinical goals and machine limits through initialization, control-point redistribution, the iterative dosimetric and sequencing/deliverability loop, and post-processing, yielding a deliverable plan with a predicted delivery time.

### *Workflow and User Interaction*

The planner specifies clinical intent by selecting a modulation level and assigning priorities to clinical metrics. The system then generates a plan and reports delivery time, modulation complexity, metric outcomes, and relevant tradeoff information. This workflow allows planners to evaluate not only whether a plan satisfies clinical goals, but also how those goals are achieved in terms of delivery time and complexity. DMAT therefore provides a practical framework for selecting an appropriate quality–time operating point for each clinical case. Because the modulation level is a single scalar input, the optimizer can also generate the full series of plans across the modulation range (−3 to +3) automatically in a single request, presenting the planner with the complete quality–time trade-off set for a given case so that the operating point best matching the treatment intent can be chosen directly rather than tuned by trial and error.

### *Modern delivery system representation for DMAT evaluation*

To evaluate the DMAT framework in the context of a machine advance with increased mechanical speed and photon beam output rates, we posit a theoretical delivery system that increases the characteristic maximum velocities, accelerations and dose rates of the Varian TrueBeam (24) by a factor of 2-2.5:

1. Maximum dose rate: 1400 MU/min → 3000 MU/min
2. Maximum gantry rotation speed for delivery: 6 deg/s → 15 deg/s
3. Maximum MLC speed:[2] 2.5 cm/s → 6.25 cm/s

### *Representative patient cases*

Three representative cases from an open access case library (prescription dose and fractionation schedule in parentheses) were used to illustrate how DMAT exposes a clinically interpretable trade-off between delivery efficiency and dosimetric refinement:

1. Head-and-neck SIB (2.0/1.8/1.6 Gy x 35 = 70/63/56 Gy)
2. Lung SBRT (10 Gy x 5 = 50 Gy)
3. Prostate SBRT SIB (10/7 Gy x 5 = 50/35 Gy)

[2] Practically, increasing the MLC speed by this amount would likely require an increased collimator elevation, so unlike TrueBeam the theoretical machine considered in this Manuscript utilizes the jawless dual layer MLC design introduced on the Halcyon platform.

For each case, solutions were produced with the same clinical goals but across discrete modulation settings from −3 to +3. Setting 0 represented an approximate baseline comparable to current VMAT implementation, whereas −1 to −3 progressively emphasized delivery efficiency (reduced modulation and shorter delivery) and +1 to +3 progressively emphasized modulation by relaxing complexity constraints and increasing control-point density. To enable a fair comparison, all seven plans for a given case were tuned so that all PTVs satisfied D95% > 95% of the applicable prescription dose, so that differences reflect modulation and delivery-time behavior rather than failure to meet the minimum target-coverage criterion. The plan quality score (Table S-1) is a composite index computed from the case-specific clinical evaluation metrics (target coverage and conformity together with the relevant organ-at-risk dose objectives), where each metric is scored against its clinical goal and the scores are summed using the weighting scheme reported in the source study (34); higher values denote better overall plan quality. The modulation complexity score (MCSv) is computed from the control-point sequence as the product of an aperture-area-variability term and a leaf-sequence-variability term aggregated over the arc and weighted by segment MU (35); values approach 1 for simple, near-open apertures and decrease with increasing modulation, so 1−MCSv is reported as an intuitive measure of modulation complexity. Additional parameters compared included total MU and MCSv.

## Results

### *Representative cases dose distributions*

Figure 3 shows the dose color wash at a representative plane for three different patient cases across seven modulation complexity settings. In all cases the level-0 setting corresponds to the VMAT-like fixed-CP reference defined above, and the negative (−1 to −3) and positive (+1 to +3) levels are presented and interpreted relative to this same-machine baseline. Delivery time and modulation surrogate MU/Gy is shown for each dose distribution. Increasing modulation generally enabled more refined dose shaping, although the magnitude and clinical relevance of the improvement were strongly case dependent; the head-and-neck case, for example, showed improved target coverage, dose homogeneity, and parotid sparing. Higher modulation settings also resulted in more spatially selective dose distributions, reflected by increasingly prominent low-to-mid dose streaking in regions contributing to the elevated modulation. Delivery time increased progressively with modulation level, directly illustrating the explicit trade-off between dosimetric quality and delivery efficiency.

At higher modulation levels, the prostate case exhibits several prominent low-to-mid-dose directions resembling a fixed-field arrangement. These are not predefined beam angles; they emerge from the case-specific dosimetric optimization and control-point redistribution based on delivery-complexity and anatomical information.

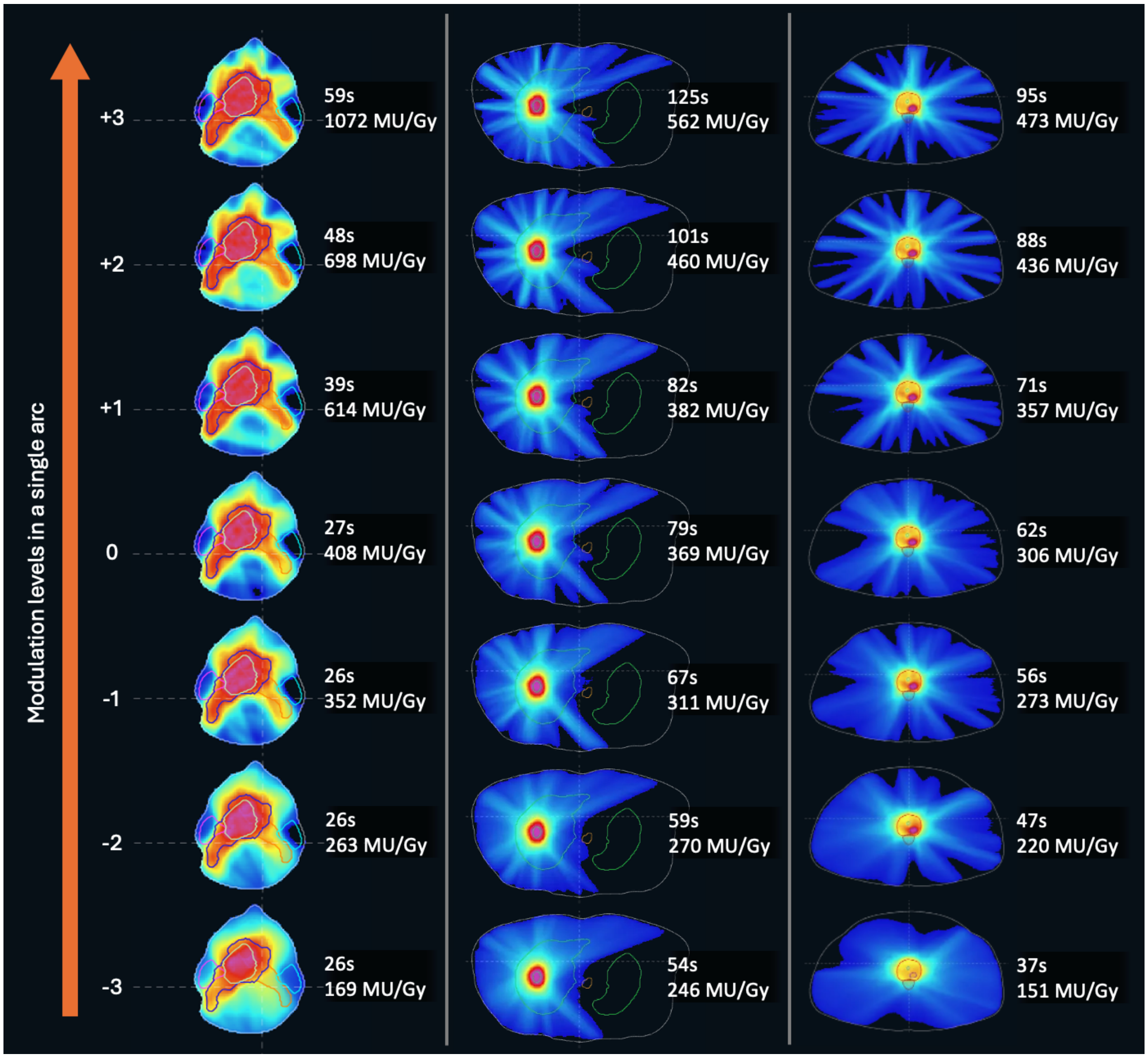

Figure 3 Representative comparison of DMAT solutions across modulation levels (−3 to +3) for three single-arc cases: head-and-neck, lung SBRT, and prostate SBRT. Rows correspond to the user-selected modulation level, with increasing modulation indicated upward. For each case, the displayed dose pattern and accompanying quantitative labels report delivery time (s) and normalized monitor-unit intensity (MU/Gy). Increasing modulation generally increases delivery time and MU/Gy, reflecting greater allocation of delivery resources to achieve more refined dose shaping; conversely, lower modulation levels favor shorter delivery with reduced modulation complexity. Together, these examples illustrate the explicit quality–time trade space exposed by the DMAT framework and its dependence on disease site.

Figure 4 plots each published clinical goal (36) against the delivery time of the corresponding DMAT solution, with markers distinguishing goals that were met, met only at the acceptable level, or not met. The three sites showed distinct patterns. For head-and-neck, attainment of the more demanding target-coverage goals improved as delivery time increased. For lung SBRT, all evaluated goals were met across the full delivery-time range, indicating little benefit from additional modulation. For prostate SBRT, goals were generally maintained across modulation levels except at the most efficiency-biased setting (−3), where GTV-boost coverage and the left femoral-head constraint failed; both recovered at less efficiency-biased levels. Together, these data help plan evaluators determine for each site a delivery-time threshold below which published goals fail and above which additional delivery time provides diminishing benefit.

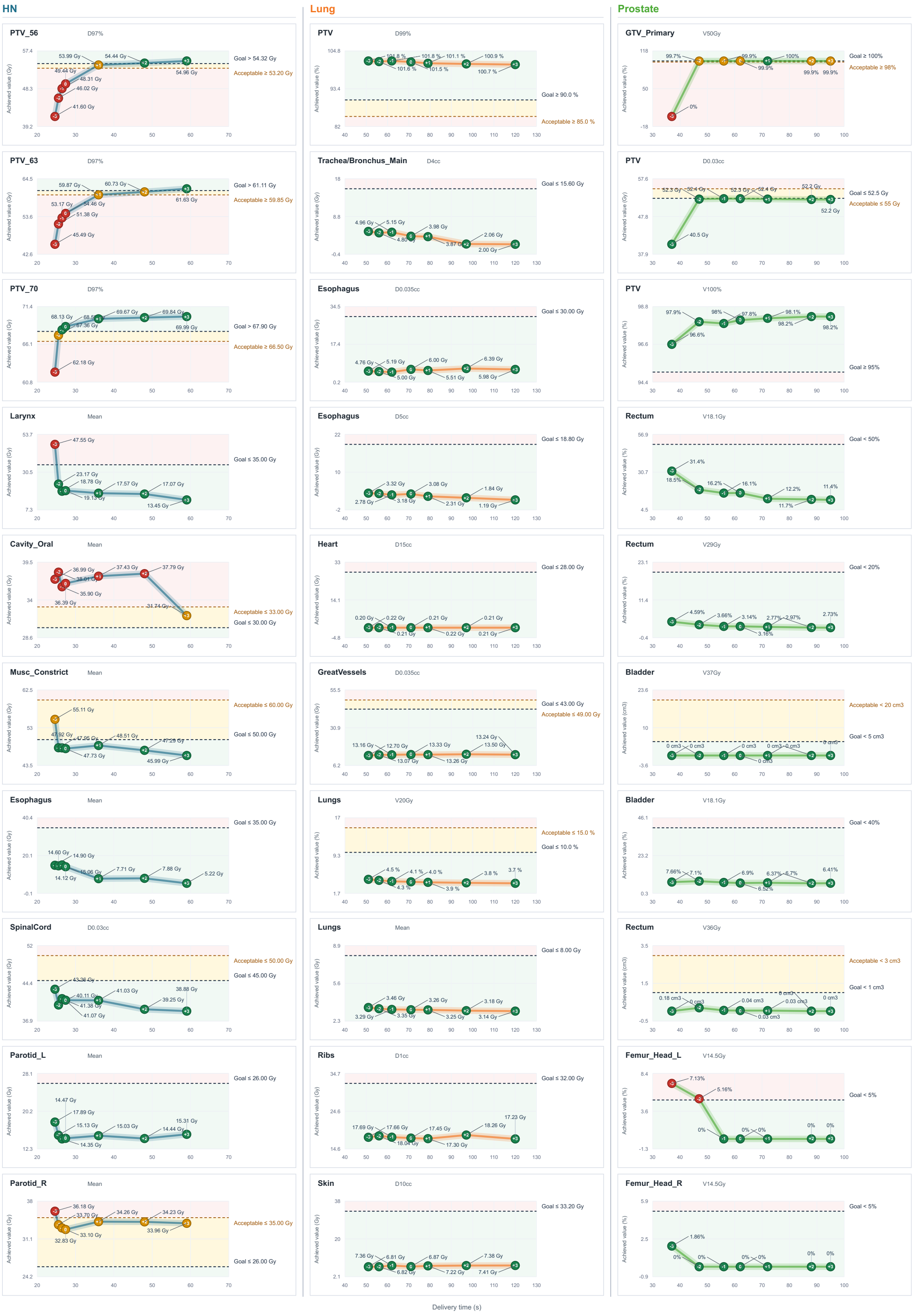

HN
Lung
Prostate
PTV_56 D97%
Goal > 54.32 Gy
Acceptable ≥ 53.20 Gy
PTV_63 D97%
Goal > 61.11 Gy
Acceptable ≥ 59.85 Gy
PTV_70 D97%
Goal > 67.90 Gy
Acceptable ≥ 66.50 Gy
Larynx Mean
Goal ≤ 35.00 Gy
Cavity_Oral Mean
Acceptable ≤ 33.00 Gy
Goal ≤ 30.00 Gy
Musc_Constrict Mean
Acceptable ≤ 60.00 Gy
Goal ≤ 50.00 Gy
Esophagus Mean
Goal ≤ 35.00 Gy
SpinalCord D0.03cc
Acceptable ≤ 50.00 Gy
Goal ≤ 45.00 Gy
Parotid_L Mean
Goal ≤ 26.00 Gy
Parotid_R Mean
Acceptable ≤ 35.00 Gy
Goal ≤ 26.00 Gy
PTV D99%
Goal ≥ 90.0 %
Acceptable ≥ 85.0 %
Trachea/Bronchus_Main D4cc
Goal ≤ 15.60 Gy
Esophagus D0.035cc
Goal ≤ 30.00 Gy
Esophagus D5cc
Goal ≤ 18.80 Gy
Heart D15cc
Goal ≤ 28.00 Gy
GreatVessels D0.035cc
Goal ≤ 43.00 Gy
Acceptable ≤ 49.00 Gy
Lungs V20Gy
Acceptable ≤ 15.0 %
Goal ≤ 10.0 %
Lungs Mean
Goal ≤ 8.00 Gy
Ribs D1cc
Goal ≤ 32.00 Gy
Skin D10cc
Goal ≤ 33.20 Gy
GTV_Primary V50Gy
Goal ≥ 100%
Acceptable ≥ 98%
PTV D0.03cc
Goal ≤ 52.5 Gy
Acceptable ≤ 55 Gy
PTV V100%
Goal ≥ 95%
Rectum V18.1Gy
Goal < 50%
Rectum V29Gy
Goal < 20%
Bladder V37Gy
Acceptable < 20 cm3
Goal < 5 cm3
Bladder V18.1Gy
Goal < 40%
Rectum V36Gy
Acceptable < 3 cm3
Goal < 1 cm3
Femur_Head_L V14.5Gy
Goal < 5%
Femur_Head_R V14.5Gy
Goal < 5%
Achieved value (Gy)
Achieved value (%)
Achieved value (cm3)
Delivery time (s)

Figure 4 Clinical goal attainment versus delivery time across all three representative sites (head-and-neck, lung SBRT, and prostate SBRT). Each marker represents one published structure/metric goal plotted against the delivery time of the corresponding DMAT solution across modulation levels −3 to +3; marker style/color denotes whether the primary goal was met, only the acceptable (secondary) level was met, or the goal was not met.

### *Variable control-point density with increasing modulation*

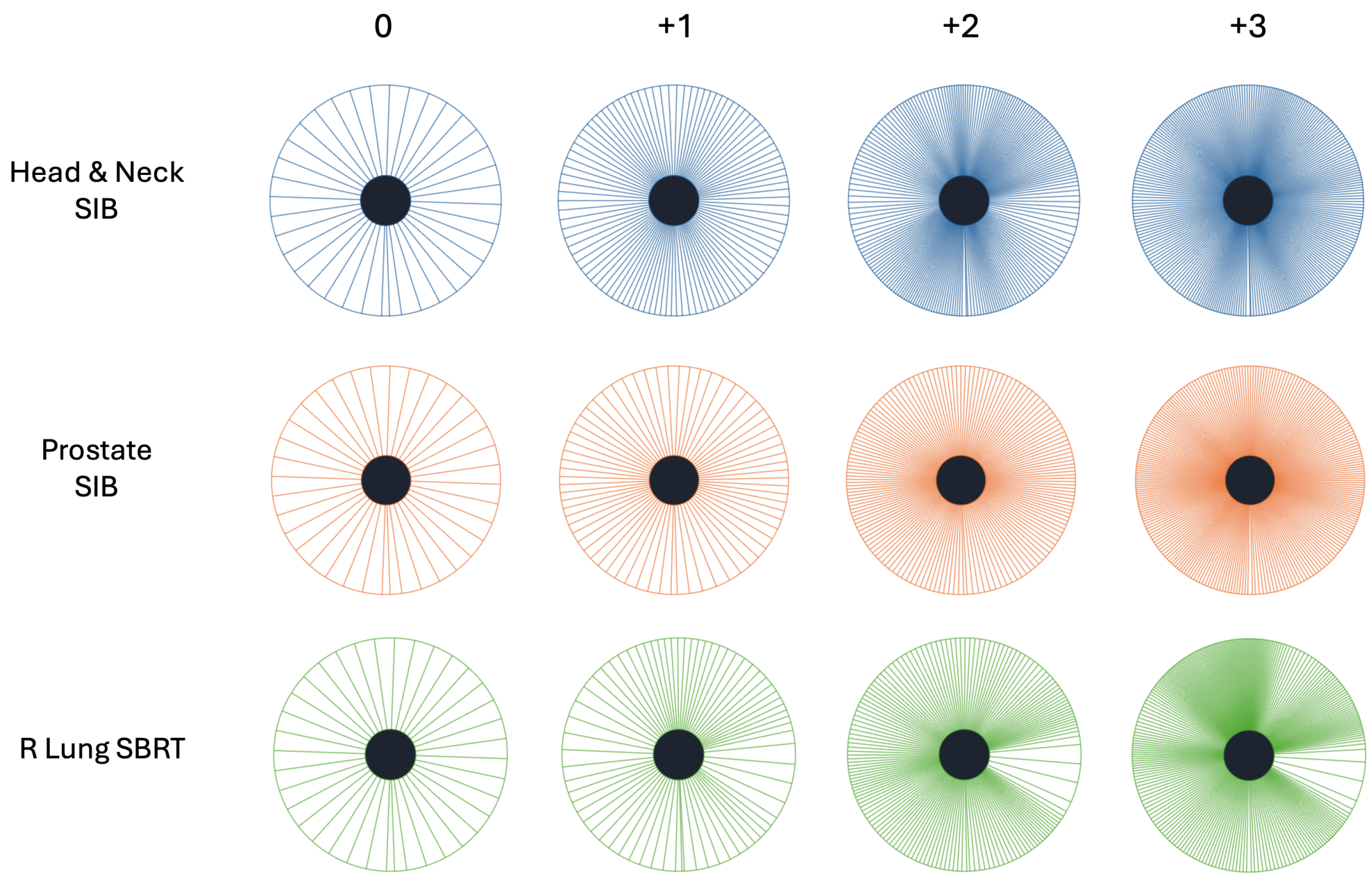

Figure 5 Illustration of variable control point density for different plans. One spoke represents 5 control points. Only 0 and +1/+2/+3 modulation levels are shown as control point density is maintained the same for lower modulation levels.

Table 1 Number of control points per full arc by modulation level for each representative case. Control-point density is held fixed at the level-0 budget of 180 control points for all efficiency-biased levels (−3 to 0) and increases progressively and case-specifically for the positive levels (see Figure 6 for the corresponding delivery-time, total-MU, and plan-quality behavior).

| Modulation Level | Head-and-Neck SIB | Lung SBRT | Prostate SBRT SIB |
|---|---|---|---|
| **−3** | 180 | 180 | 180 |
| **−2** | 180 | 180 | 180 |
| **−1** | 180 | 180 | 180 |
| **0 (VMAT-like reference)** | 180 | 180 | 180 |
| **+1** | 491 | 338 | 337 |
| **+2** | 1034 | 785 | 790 |
| **+3** | 1328 | 1261 | 1241 |

Figure 5 and Table 1 summarize how DMAT adapts control-point density according to the user-selected modulation level. At levels −3 to 0, CP density is held constant at the level-0 budget of 180 control points so that efficiency-first solutions reduce complexity through tighter modulation controls (e.g., reduced leaf-travel allowance and other complexity penalties) rather than by reducing the number of control points or increasing their angular spacing. At positive modulation levels, CP density increases progressively and nonuniformly, allocating additional control points to arc sectors where greater angular resolution is expected to improve plan quality; the case-specific totals are reported in Table 1.

## *Disease-specific quality–time trade-offs*

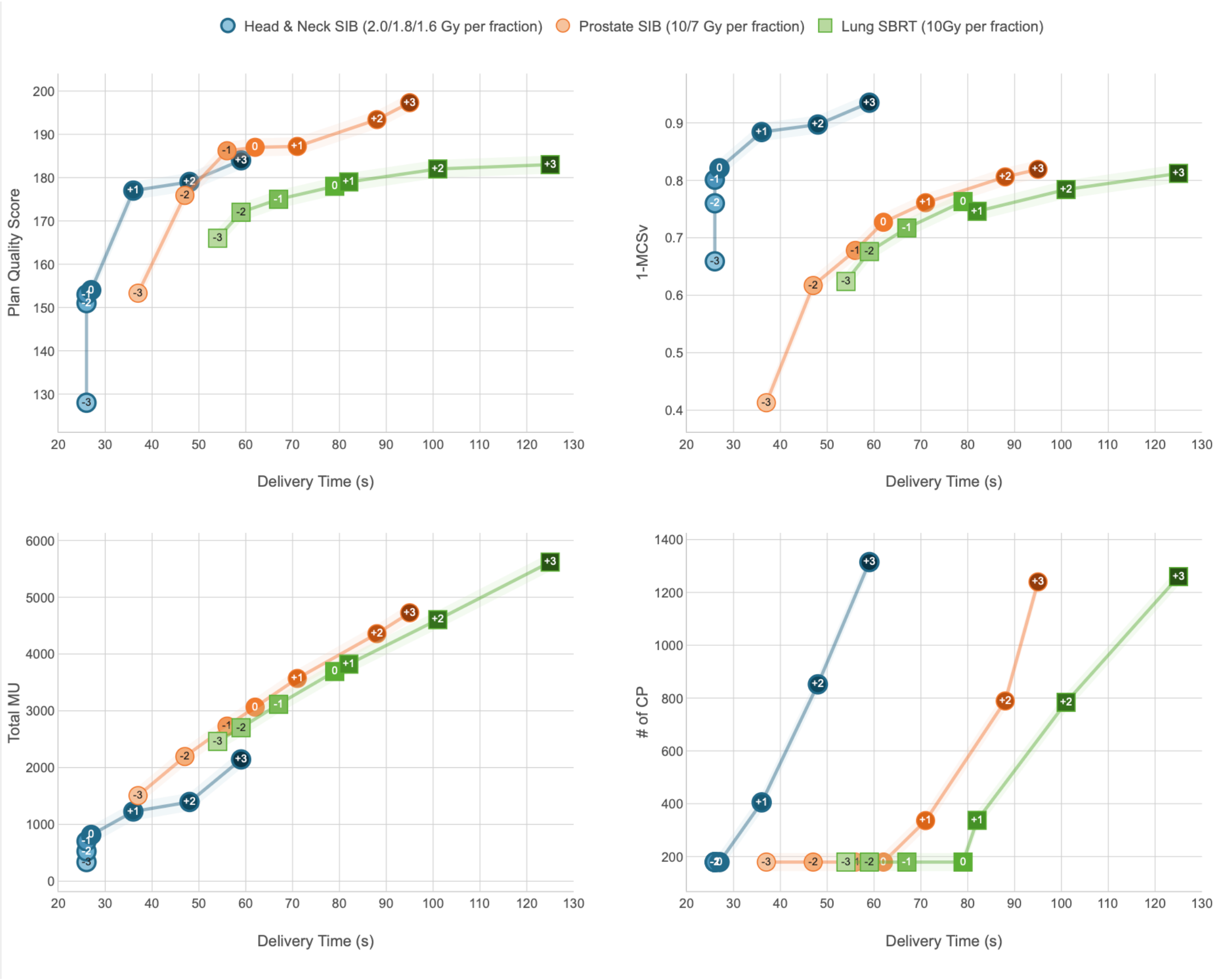

Figure 6 Delivery time versus representative plan-level performance metrics for head-and-neck (2.0/1.8/1.6 Gy per fraction), prostate SIB (10/7 Gy per fraction) and lung SBRT (10 Gy per

fraction) cases across the evaluated DMAT solutions. Each connected series traces the progression of modulation settings within a given disease site, and the primary interpretation should therefore be made within each site rather than across sites. The reported plan quality score is a composite score derived from the clinically relevant dosimetric evaluation metrics used for each case. Across modulation settings within an individual disease site, longer delivery times are generally associated with higher plan-quality scores together with increased modulation- and delivery-related quantities in the lower panels. The per-fraction dose is provided in parentheses for reference.

The relationship between plan quality and delivery time differed substantially by disease site (Figure 6). For the head-and-neck case, increasing modulation produced a coupled rise in plan quality score, delivery time, modulation complexity (1−MCSv), total MU, and number of control points. This consistent co-increase indicates that the optimizer actively exploits additional temporal degrees of freedom and more variable aperture shapes to achieve clinically meaningful improvements in target coverage and OAR sparing. Conversely, aggressively reducing modulation degraded plan quality while reducing aperture/MU modulation and delivery time at the same fixed 180-CP budget — demonstrating a disease-specific lower bound on acceptable complexity.

For prostate SBRT with SIB, increasing modulation above baseline produced smaller incremental improvements in plan quality than in head-and-neck. Nevertheless, Figure 6 shows the expected monotonic rise in delivery time with increasing modulation complexity—higher 1−MCSv, higher total MU, and increased CP count—indicating that the added delivery resources are primarily spent on additional modulation rather than unlocking large dosimetric gains. As a result, delivery efficiency can often be prioritized for this site with more predictable and clinically modest trade-offs in plan quality.

For lung SBRT, the plan-quality score varied relatively little across modulation settings compared with head-and-neck. Even as delivery time increased with higher modulation, the accompanying increases in 1−MCSv, total MU, and CP count translated into comparatively modest quality gains, consistent with a simpler geometric dose-shaping task for this case. This suggests that more efficient presets (e.g., −1 or −2) may be sufficient for similar lung SBRT geometries when the clinical objective is to minimize treatment time without materially sacrificing plan quality.

For the lung SBRT case, the increase in control-point count and total MU at higher modulation levels was not matched by a proportional gain in plan quality. This behavior is consistent with the role of the modulation level as a resource allowance rather than a quality target: increasing the level relaxes complexity penalties and permits additional control points and MU, so the optimizer is free to expend these added degrees of freedom even when their marginal dosimetric benefit is small. For a geometrically simple, low-OAR target such as this case, the additional modulation yields little incremental benefit, whereas for the prostate case a comparable increase in MU accompanies a more pronounced improvement in plan quality. This contrast reinforces the central premise of DMAT—that the appropriate operating point is anatomy dependent—and indicates that efficiency-biased presets (e.g., −1 to −3) reach equivalent quality more economically for such simple geometries. Automated adaptation of the complexity penalty to suppress non-beneficial modulation in low-complexity cases is a natural refinement of this behavior.

Examining the complexity surrogates directly reinforces this site-dependent picture. 1−MCSv scaled near-linearly with delivery time for head-and-neck, reflecting the optimizer's deliberate use of smaller and more variable apertures to resolve the interlocked PTV/OAR topology, whereas the complexity–time slope was progressively shallower for prostate SBRT/SIB and shallowest for lung SBRT, indicating that aperture fragmentation beyond a site-specific threshold contributes little additional dosimetric value. Total MU increased monotonically within each case cohort for the same prescription dose, confirming increased modulation.

CP count remained at 180 from levels −3 through 0, then rose steeply from +1 to +3, with corresponding increases in delivery time. The case-specific CP counts at positive levels demonstrate anatomy- and objective-dependent allocation.

These examples illustrate the central claim of DMAT: the optimal balance between plan quality and delivery time is not universal but depends on anatomy, disease site, and clinical intent, and should therefore be made visible during planning. This is further confirmed by the three modulation complexity surrogates: head and neck is quality-limited and rewards added complexity, prostate SBRT with SIB is modulation-saturated above baseline, and lung SBRT is time-limited, with efficient presets sufficient to meet clinical objectives.

### *Computational performance*

All dose calculation and optimization steps were performed on a single NVIDIA L4 GPU. Figure 7 shows the dose-calculation and optimization times together with the total runtime for the representative plans. Mean total runtime ranged from approximately 109.8 s (prostate SBRT) to 134.0 s (head-and-neck) across sites, comprising a dose-calculation component of roughly 7.8–13.2 s and an optimization component of roughly 96.6–126.2 s. GPU acceleration of both the dose calculation and the optimization solver produced runtimes that may support time-constrained and adaptive workflows. Runtimes are single measurements obtained with prototype research code and are governed primarily by the number of iterations required to reach the convergence criteria, which depends on the initial solution and on case-specific cost-function behavior; they should therefore be interpreted as representative prototype runtimes rather than as a systematic trend with modulation level.

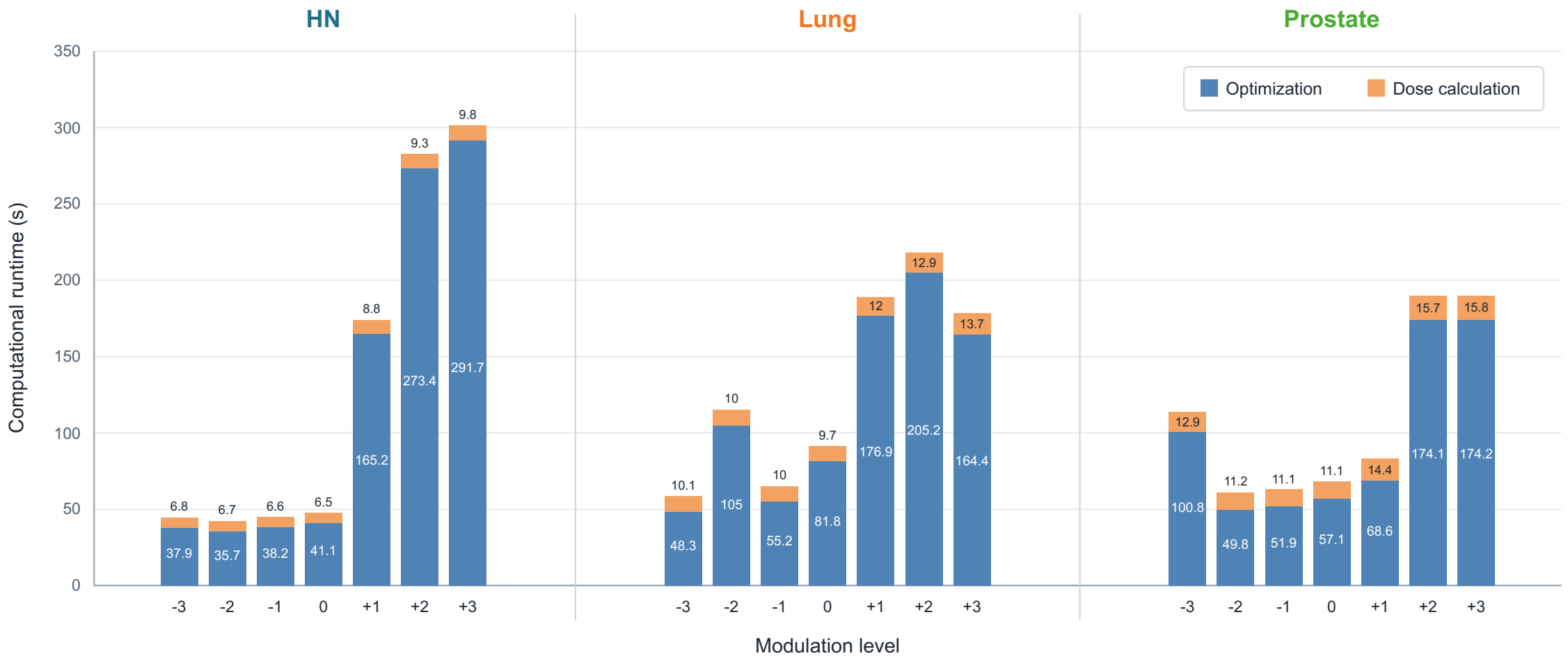

Figure 7 Computational runtime for the representative plans on a single NVIDIA L4 GPU, shown by site (head-and-neck, lung SBRT, and prostate SBRT) and modulation levels. Bars/points show the optimization and dose-calculation components and the total runtime.

## Discussion

This study introduces DMAT as a time-aware, modulation-steered framework that jointly considers plan quality, delivery time, modulation complexity, and machine-specific delivery characteristics. Across the representative cases, the value of additional modulation was anatomy dependent: the head-and-neck case showed substantial improvements with increasing modulation, whereas prostate and lung SBRT showed smaller gains beyond baseline. These findings illustrate the value of exposing the quality–time trade-off during planning rather than treating delivery time and complexity as emergent plan properties.

All comparisons in this study are made relative to the level-0 VMAT-like fixed-control-point reference generated using the same prototype optimizer and emulated machine, not against commercial VMAT. Because the accelerated machine is a research construct that cannot be planned using the commercial Photon Optimizer, this study characterizes modulation steering within the DMAT framework rather than establishing superiority over VMAT on current clinical systems. DMAT is an arc plan-generation framework defined by machine emulation and dynamic, time- and complexity-aware optimization; the hypothetical accelerated machine serves only as a demonstration platform. DMAT models gantry, MLC, and dose-rate dynamics jointly, with the rate-limiting axis depending on the machine, prescription, and requested modulation.

The principal contribution of DMAT is therefore a time-aware, navigable optimization framework rather than a fixed dosimetric advantage over VMAT. Each modulation level represents a distinct operating point: additional delivery resources can be converted into greater modulation where clinically beneficial, whereas efficiency-biased levels preserve comparable quality at shorter delivery times when added complexity provides little return. Adaptive control-point allocation concentrates angular resolution where it is expected to be most useful. DMAT thereby makes delivery time a calculated and steerable planning quantity rather than an emergent result of a fixed control-point representation (5–8,25). Leaf travel, monitor units, aperture complexity, and control-point placement are optimized together against an explicit model of the machine. The modulation that can be achieved along a given trajectory is therefore limited by what the machine is mechanically capable of, rather than by control-point settings fixed in advance, and the high axis speeds and dose rate of the emulated delivery system are an enabling condition in this respect. In conventional VMAT, how much the aperture shape can change within a single rotation is itself a limit, and the modulation that complex cases require is usually obtained by adding overlapping arcs over the same angular range. Reaching the plan quality reported here within a single arc therefore reflects both the joint optimization of apertures and control-point placement and the mechanical capability of the delivery system; neither alone would be sufficient. Single arcs were used throughout this study as a proof of concept, and the framework applies equally to multi-arc and noncoplanar trajectories.

The composite plan-quality score summarizes the overall quality–time trend but does not replace review of individual clinical metrics. It is most useful for judging how much overall quality is gained or lost between modulation levels; because it is a weighted aggregate, however, the highest-scoring solution is not necessarily the best for any particular dosimetric endpoint. Evaluators should therefore examine the specific endpoint of interest against delivery time (Figure 4) when selecting an operating point. Delivery-time efficiency is therefore offered as a selectable trade-off rather than a fixed priority over normal-tissue sparing.

Reporting delivery time alongside plan quality is not intended to assert that differences of seconds are clinically decisive, but to make navigable a dimension of the planning problem that previously was not. Conventionally a plan is judged on dosimetric quality alone, and its delivery time is whatever the chosen technique, arc count, and control-point sampling happen to produce — known only after planning and altered only by re-planning. Because delivery time is computed within the optimizer and the full modulation series follows from a single planning request, quality and delivery time are instead presented together as a continuum, and a point on it can be selected deliberately for the case at hand. Whether, and by how much, movement along this axis translates into clinical benefit remains an open question for studies with clinical endpoints. Independently of that question, the same emulation that yields the time estimate keeps the optimized trajectory within the dynamic limits of the machine, so that the plan that is scored is the plan that can be executed.

Outside of SRS, treatment-session time is usually dominated by setup, imaging, registration, and review. DMAT therefore treats delivery time as a case- and machine-specific trade-off rather than an objective to minimize indiscriminately. Shorter delivery time is most valuable in time-sensitive settings such as breath-hold delivery and intrafraction-motion management (30,37–39). Shorter beam-on time may additionally reduce dose to circulating blood, although the magnitude of any clinical effect and the relevance of a specific time threshold remain uncertain (40–43).

Beyond these immediate applications, the same time-as-a-resource framework may support broader clinical and operational benefits. Integration of automated clinical-goal generation, machine-aware timing, dose calculation, and optimization could allow multiple modulation settings or alternative trade-off solutions to be evaluated within time-constrained adaptive workflows. When combined with advances in image guidance (44–46), this capability may reduce on-couch planning latency and expand the solution space that can be considered during adaptation. More predictable and efficient delivery may also improve treatment capacity and facilitate complex paradigms that are currently limited by delivery time, including polymetastatic treatments using fewer isocenters. At the clinic level, making the quality–time trade space visible could support more consistent plan selection, reduce trial-and-error refinement, and help translate advances in machine speed, imaging, and automation into scalable treatment workflows. These potential benefits require prospective evaluation and should not be interpreted as demonstrated clinical or operational outcomes.

A central strength of DMAT is unification – the integration of concepts previously studied largely as separate components, including nonuniform control-point spacing, dynamic trajectories, multicriteria optimization, machine-aware timing, and direct clinical-metric optimization(15–19,22,23,25). Unlike approaches focused primarily on angular supersampling, DMAT couples control-point placement with machine emulation and coordinated control of leaf travel, MU, aperture complexity, and delivery time. The contribution is therefore not control-point redistribution alone but a unified framework in which quality, time, complexity, and deliverability are optimized together.

The appropriate quality–time operating point is unlikely to be universal. It will depend on anatomy, organ-at-risk proximity, motion characteristics, fractionation, machine capabilities, and workflow constraints. Presenting the available operating points explicitly may help planners apply complexity where it provides measurable benefit while preserving efficiency where it does not.

This study has several limitations. It is a proof-of-concept planning evaluation based on three representative cases, a hypothetical accelerated machine, and prototype implementations that do not represent current commercial product capabilities. The results are limited to modeled delivery time and dosimetric surrogates and do not include end-to-end delivery validation on a clinical machine, patient-specific QA, or clinical outcomes. Because DMAT is machine-parameterized rather than machine-agnostic, implementation on a clinical linac would require time-resolved characterization and commissioning of machine dynamics, axis synchronization, controller behavior, trajectory smoothing, and delivery tolerances. The figures should therefore be interpreted as illustrative examples of the quality–time trade space rather than site-wide benchmarks or treatment prescriptions. Future work will include measurement-based validation, including log-file–based characterization of how often, and by how much, acceleration limits extend delivery on a clinical system and how closely predicted and delivered times agree, development of disease-specific acceptance criteria, evaluation of whether shorter delivery translates into measurable clinical or operational benefit, extension to additional complex cases, including paraspinal treatments, comparison of collimation designs such as dual-layer, high-definition, and jaw-tracking configurations within the same framework, automated adaptation of the complexity penalty to limit non-beneficial modulation, and automated identification of discrete high-modulation sectors within an otherwise continuous arc.

## Conclusions

DMAT introduces a time-aware, modulation-steered radiotherapy planning paradigm that explicitly couples plan quality, modulation complexity, and machine-specific delivery characteristics. By embedding machine-aware timing, adaptive control-point allocation, and clinically meaningful metric-based optimization, DMAT makes the quality–time trade-off visible and navigable, allowing delivery resources to be allocated where they provide measurable benefit while preserving efficiency where they do not. The result is a practical and interpretable foundation for next-generation radiotherapy solutions, particularly in settings where motion management, adaptive readiness, and scalable high-quality treatment are critical.

## Acknowledgments

The authors thank colleagues at Varian Medical Systems, Inc., a Siemens Healthineers company, for technical discussions and support during development of this work.

## Ethics Statement

This study did not involve human or animal subjects. All work was performed as a treatment-planning study using de-identified, publicly available representative cases; institutional review board approval and informed consent were therefore not required.

## Author Contributions

Taoran Li and Esa Kuusela: conceptualization, methodology, data, investigation, writing – original draft. Esa Kuusela, Emmi Ruokokoski, Heini Hyvönen, Mirko Myllykoski, Jussi Nurminen, Riku Paananen, Jarkko Peltola, Marko Rusanen, and Martin Sabel: methodology, technical development, review and editing. Jerry Jaboin, Kevin Moore, and Christopher Boylan: supervision, review and editing. All authors reviewed and approved the final manuscript.

## Data Availability Statement

Data presented in this manuscript were based on representative treatment-planning cases publicly available at https://cancermedaffairs.siemens-healthineers.com/probeam-case-studies.

## Conflict of Interest

All authors are employees of Varian Medical Systems Inc., a Siemens Healthineers company.

## Funding

This work was supported by Varian Medical Systems Inc., a Siemens Healthineers company. No external funding was received.